\documentclass{elsart}
\usepackage{graphicx}
\usepackage{amsmath}
\usepackage{amssymb}

\journal{Journal of Alloys and Compounds}

\begin{document}

\begin{frontmatter}

\title{YbGaGe: normal thermal expansion}

\author[AL]{Y. Janssen\corauthref{cor}}
\corauth[cor]{Corresponding author.} \ead{yjanssen@ameslab.gov}
\author[AL]{S. Chang}
\author[AL,GIST]{B.K. Cho}
\author[LANL]{A. Llobet}
\author[AL]{K.W. Dennis}
\author[AL]{R.W. McCallum}
\author[AL,ISU]{R.J. Mc Queeney}
\author[AL,ISU]{P.C. Canfield}

\address[AL]{Ames Laboratory, 50011 Ames IA}
\address[GIST]{Center for Frontier Materials and Department of
Materials Science and Engineering, GIST, Kwangju 500-712, South
Korea}
\address[LANL]{Los Alamos National Laboratory, 90120 Los Alamos, NM}
\address[ISU]{Department of Physics and Astronomy, Iowa State University, 50011 Ames IA}

\begin{abstract}

We report evidence of the absence of zero thermal expansion in
well-characterized high-quality polycrystalline samples of YbGaGe.
High-quality samples of YbGaGe were produced from high-purity
starting elements and were extensively characterized using x-ray
powder diffraction, differential thermal analysis, atomic emission
spectroscopy, magnetization, and neutron powder diffraction at
various temperatures.  Our sample melts congruently at
920$^{\circ}$~C. A small amount of Yb$_2$O$_3$ was found in our
sample, which explains the behavior of the magnetic
susceptibility. These observations rule out the scenario of
electronic valence driven thermal expansion in YbGaGe. Our studies
indicate that the thermal expansion of YbGaGe is comparable to
that of Cu.
\end{abstract}

\begin{keyword} YbGaGe, intermetallics, neutron diffraction,
magnetic measurements, thermal analysis
\end{keyword}

\end{frontmatter}

\section{Introduction}

Recently, it was reported that the compound YbGaGe shows
negligible thermal expansion between 100 and
400~K~\cite{Salvador}.  This effect was attributed to a
progressive change in the Yb electronic valence. It was also noted
that the exact composition is crucial, since for some of the
YbGa$_{1+x}$Ge$_{1-x}$ the zero thermal expansion occurs, whereas
negative and positive thermal expansion also occur, presumably for
other values of $x$. However, the precise values for {\em x} were
not reported. In a more recent publication by the same
authors~\cite{Margadonna04}, the compound with composition
YbGa$_{1.05}$Ge$_{0.95}$ was reported to have a sudden valence
transition at 5 K, accompanied by a large change in the lattice
parameters, as determined by high-resolution X-ray powder
diffraction.

Most materials having negative thermal expansion (NTE) are
insulating oxides, where the NTE effect is caused by an
underconstrained crystal lattice~\cite{[ref1]}.  The number of
metals possessing NTE are few, with the most notable case being
the Invar alloys.  In the metallic NTE materials, usually two
local atomic configurations with different atomic sizes exist,
with the smaller sized atom being an excited state.  These atomic
configurations can be different magnetic (Invar) and/or valence
states. Delta-plutonium is a possible example of an NTE material
caused by a valence transition and has received considerable
attention lately~\cite{[ref2]}. However, Pu is difficult to study,
making YbGaGe a potentially very interesting material for the
study of NTE behavior of electronic origin.

We have prepared single-phase polycrystalline YbGaGe by combining
high-purity Yb, Ga and Ge, and characterized it by means of x-ray
powder diffraction, chemical analysis, differential thermal
analysis and measurements of magnetization. We also performed
neutron powder diffraction in order to measure the temperature
dependence of the lattice parameters, to further check the quality
of the sample and to determine the distribution of the Ga and Ge
ions within the lattice. Neutron diffraction was employed since Ga
and Ge are nearly indistinguishable to standard x-ray diffraction
techniques.

\section{Sample synthesis}

Salvador et al.~\cite{Salvador} reported that they synthesized
YbGaGe in the following, single-step, manner. They mention that
the elements were directly combined in their stoichiometric ratios
and heated to 850$^\circ$~C for 96 hours followed by a cool to
room temperature over about 12 hours. The authors did not indicate
the purity of their starting materials, nor did they mention the
material of the crucibles that they used.

For the synthesis of our sample we used the high-purity elements
Yb (Ames Laboratory, 99.94\% elements basis), Ga (Alfa Aesar,
99.999\% metals basis) and Ge (Alfa Aesar, 99.999\% metals basis).
It is important to use elements of high-purity, because it is well
known that impurities can stabilize or even enable the formation
of intermetallic phases. Stoichiometric amounts, of liquid Ga and
lumps of Ge and Yb, with a total mass of about 8~g, were put in a
Ta crucible.  We have found Ta crucibles to be inert for use with
Yb, Ga and Ge in the temperature range and composition of
interest. To prevent evaporation of the volatile Yb, the crucible
was sealed by arc welding in an Ar atmosphere. The Ta crucible was
placed in a silica ampoule that was evacuated and subsequently
sealed.

The ampoule was placed in a furnace at room temperature. The furnace
was heated to 200$^\circ$~C in 2 hours and, after that, to
850$^\circ$~C in 70 hours. The temperature was kept constant at
850$^\circ$~C for 40 hours, after which it was lowered to room
temperature in 24 hours.

The sample came out of the Ta crucible easily and was brittle.
However, the shapes of the Yb pieces were still discernible,
suggesting that the formation of the desired phase was incomplete.
A small amount, less than 0.5~g, of the sample was finely ground
and used for X-ray powder diffraction. The strongest reflections
on the diffractogram could be indexed according to the reported
crystal structure of YbGaGe~\cite{Salvador}, however, reflections
belonging to undetermined secondary phases were clearly visible.

The remainder of the sample was coarsely ground and mixed for
homogenization, and pressed into two pellets. These pellets were
sealed in a new Ta crucible, that was sealed and put in a silica
ampoule as described above. The ampoule was heated in a furnace up to
860$^\circ$~C, and kept at this temperature for 10 days. After this,
the ampoule was taken out of the furnace and allowed to cool down
naturally.

After this second heat treatment, the pellets had their original
shape, but the surfaces looked as though they had partially
melted. This indicates the presence of a metallic liquid during
formation. The sample was again brittle. X-ray powder diffraction
on different parts of the sample did not show any trace of any
other phase besides YbGaGe. This X-ray single-phase sample was
used for the experiments described below.

\section{Magnetization and thermal properties}

The temperature dependent magnetization of YbGaGe was measured in
a Quantum Design Magnetic Properties Measurement System on a bulk
piece of approximately 300 mg. Particular care was taken in the
measurement of this low-moment sample, as extremely small
quantities of ferromagnetic impurities or the sample holder may
produce a small zero offset which can be independent of
temperature.  If no correction is made, an offset will give the
appearance of large deviations from Curie-Weiss behavior at high
temperatures, suggesting a temperature dependent magnetic moment.
Therefore measurements were made in applied fields of 0.1 and 1.0
T and the differential susceptibility was calculated eliminating
the effects of any constant offset.

\begin{figure}[!tb]
\begin{center}
  \includegraphics[width=\textwidth]{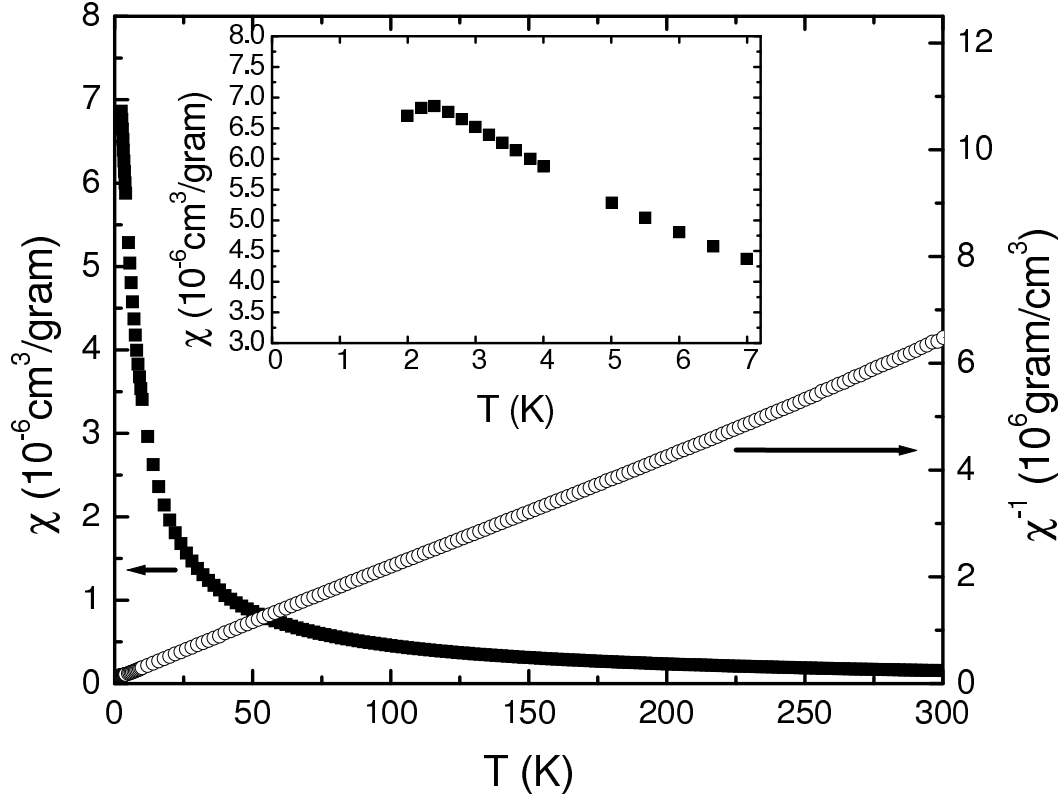}
\caption{Magnetic gram-susceptibility (right) and reciprocal
  gram-susceptibility (left) of our YbGaGe sample. The inset is an
  enlargement of the susceptibility in the low-temperature
  region.}\label{susceptibility}
\end{center}
\end{figure}

 The temperature dependent differential magnetic susceptibility is displayed in
Figure~\ref{susceptibility}. It shows a maximum near about 2.3 K (see
inset), and decreases continuously as a function of temperature.
Figure~\ref{susceptibility} also shows the temperature dependent
reciprocal susceptibility. Above $\sim 5$~K, it appears as a nearly
straight line, with a slope of about $21 \times 10^3$~g~cm$^{^-3}$~K$^{-1}$.

The maximum in susceptibility near 2.3~K is due to the presence of
Yb$_2$O$_3$ in the sample, which has been reported to order
antiferromagnetically at that temperature (See e.g.
Ref.\cite{Li94}). In this framework, the Curie-Weiss behavior of
the reciprocal susceptibility determined for this sample may be
interpreted as being solely due to about 0.4wt\% of an Yb$_2$O$_3$
impurity present in the sample. This level of impurity is well
below the detection level for standard X-ray diffraction. The
origin of the oxygen in our sample is very likely associated with
the intermediate grinding of the sample, in the form of adsorbed
gas. It will be shown below that this is consistent with elemental
analysis and neutron diffraction.

Our results are inconsistent with those reported by Salvador et
al.\cite{Salvador}. They are consistent with those reported by
Bobev et al.~\cite{Bobev04}, who ascribe the susceptibility of
their sample to be mainly due to an impurity phase, and by Muro et
al.~\cite{Muro04}, who observed a weak temperature-independent
diamagnetic signal (about $-4.6\cdot 10^{-5}$ emu/mol) from their
sample. It should be noted, that our sample does not show a
temperature-independent susceptibility; an estimate of a
temperature-independent term of the susceptibility, $\chi_0$, from
a plot (not shown here) of $\chi$ vs. $1/T$ in the
high-temperature limit yielded a value that is indistinguishable
from zero.

Atomic emission spectroscopy on our sample was performed by the
commercial laboratory NSL Analytical Services Inc.  The Yb-Ga-Ge
ratio was found to be stoichiometric to within the experimental
uncertainties of the analysis (on average ~2\% for each element).
A small amount of O was identified in our sample. Assuming that
all the O atoms belong to an Yb$_2$O$_3$ impurity, we can assign
approximately 0.59(8)wt.\% of our sample to the impurity phase,
consistent with our magnetization results.

\begin{figure}[!tb]
\begin{center}
  \includegraphics[width=\textwidth]{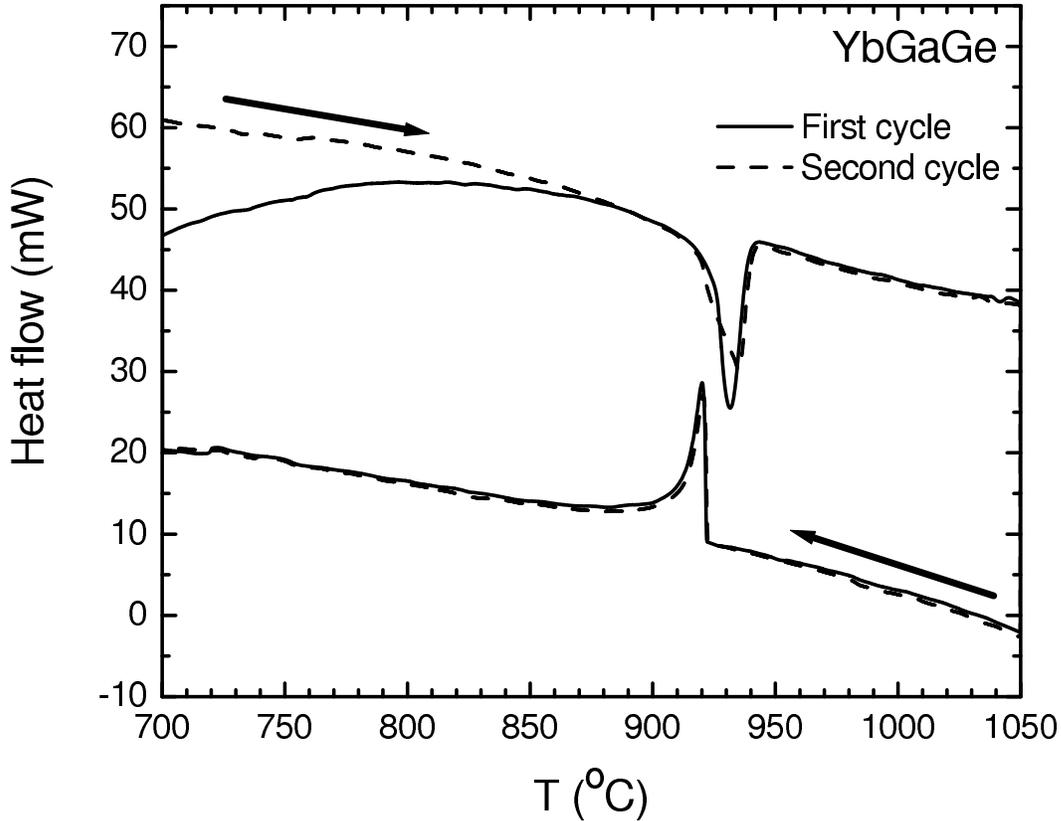}
\caption{Relevant part of the DTA trace of our sample of YbGaGe
  measured upon heating and cooling with a 10$^\circ$C/min rate. The
  first heating/cooling cycle is indicated by a full line, the second
  by a dashed line.}\label{DTAfig}
\end{center}
\end{figure}

Differential thermal analysis (DTA) was performed in a PerkinElmer
Pyris DTA 7 differential thermal analyzer. Our instrument has been
set up to provide an extremely low partial pressure of oxygen
during the measurement. The partial pressure of oxygen of the
Zr-gettered ultra-high-purity Ar process gas was monitored at the
output of the DTA furnace, and was less than 250 ppm during the
measurement. Initial measurements performed in an Al$_2$O$_3$
crucible showed clear evidence that the sample, after melting,
reacted with the crucible. In order to avoid this problem, a
small-diameter Ta tube was used to fabricate a crucible, that was
put inside the standard Al$_2$O$_3$ crucible. Figure~\ref{DTAfig}
displays the DTA results obtained using the Ta crucible. The
sample was cycled between 600$^\circ$~C to 1160$^\circ$~C twice,
at a rate of 10$^\circ$~C/min. Upon heating, a sharp endotherm
with an onset temperature of about 920$^\circ$~C is observed. No
further thermal events were observed up to 1160$^\circ$~C. Upon
cooling, a corresponding exotherm is observed. This reproducible
behavior indicates that we have eliminated both a possible
oxidation due to the process gas and a reaction with the crucible.
Furthermore, our results are consistent with YbGaGe congruently
melting around 920~$^\circ$~C. This hypothesis was verified by
heating appropriate amounts of the constituent elements  up to
1190~$^\circ$~C in a sealed Ta crucible and subsequently cooling
to room temperature in about 12 hours. Powder X-ray diffraction
indicated this sample to be single phase YbGaGe.

The results of our differential thermal analysis are inconsistent with
those reported by Salvador et al.~\cite{Salvador}, who saw no melting
or phase change up to 1000$^\circ$~C. The congruent
melting is consistent with Bridgman growth out of a stoichiometric
melt as reported by Muro et al~\cite{Muro04}.

\section{Neutron diffraction}

Time-of-flight neutron powder diffraction data were collected at
12 different temperatures between 20~K and 300~K for about 3 hours
at each temperature on the High Intensity Powder Diffractometer at
the Los Alamos Neutron Science Center.  The experiments were
performed on 3.20~g of powdered YbGaGe sealed in a vanadium can.
Temperature control was provided by a closed-cycle He
refrigerator.  Data from the $\pm 153^\circ$ banks were
simultaneously analyzed with the Rietveld refinement package
GSAS~\cite{GSAS}.

Analysis of our neutron-powder-diffraction data confirmed that our
sample contains a majority of well crystallized YbGaGe phase and
about 0.5wt\% of Yb$_2$O$_3$. This is consistent with
our atomic emission spectroscopy results, and provides further
justification for our approach to understanding the susceptibility of
our sample.

The Bragg peaks from the main phase were consistent with the
hexagonal structure (spacegroup $P6_3/mmc$, number 194) reported
in Ref.~\cite{Salvador} for YbGaGe. Therefore, in our refinements
we have assumed the Yb atoms to occupy the reported $2a$ and $2b$
sites, whereas the Ga and Ge atoms occupy two different $4f$
sites. We used the 20~K data to test various models which allow
different arrangements of the Ga and Ge atoms on the two 4$f$
sites, ranging from a totally random distribution to a completely
ordered distribution.  The best fit to the observed diffraction
profiles were given by a model with an ordered distribution of Ga
and Ge atoms, in agreement with the assignments made in
Ref.~\cite{Salvador}. However, we do find a small improvement in
the refinement if we allow  a Ga deficiency of about 6.4(6)at\%.
The improvement is marginal and inconsistent with the atomic
emmission spectroscopy results. The refined structural parameters,
as well as the lattice parameters, at 20~K are shown in
Table~\ref{20K struct para}.

\begin{table}[!tb]
\caption{Refined structural parameters of YbGaGe at 20~K. For this
fit $\chi^2=4.778$, and the weighted residual value $R_{wp} =
2.47\%$.
 }
\label{20K struct para}
\begin{center}
\begin{tabular}{c c c c c c}
\hline \hline
atom & site & occupation & x & y & z \\
\hline
Yb$_1$ & $2a$ & 1 & 0 & 0 & 0\\
Yb$_2$ & $2b$ & 1 & 0 & 0 & 1/4\\
Ga & $4f$ & 1 & 1/3 & 2/3 & 0.15403(3)\\
Ge & $4f$ & 1 & 1/3 & 1/3 & 0.61312(4)\\
\hline
\multicolumn{6}{l}{Spacegroup: $P6_3/mmc$}\\
\multicolumn{6}{l}{$a = 4.19186(2)$}\\
\multicolumn{6}{l}{$c = 16.6709(2)$}\\
\end{tabular}
\end{center}
\end{table}

\begin{figure}[!tb]
\begin{center}
\includegraphics[width=\textwidth]{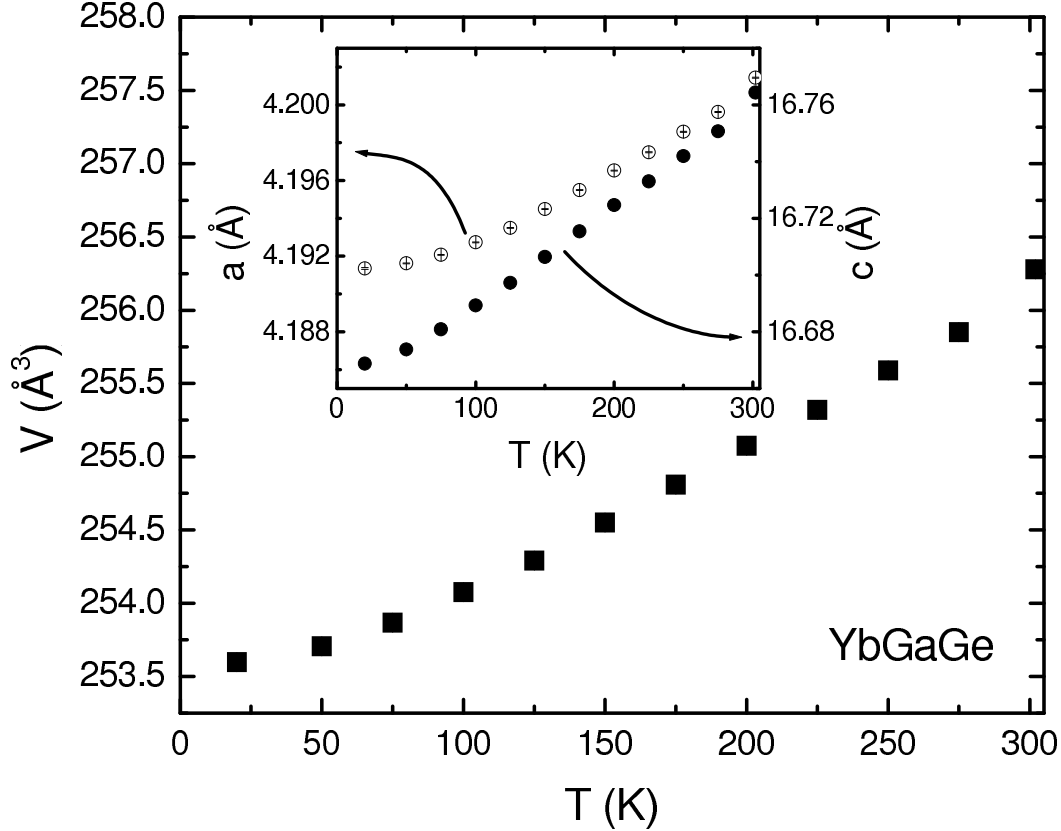}
\caption{Unit cell volume of YbGaGe as a function of temperature
obtained from neutron diffraction experiments.  The inset shows
the the lattice parameters $a$ (open circles) and $c$ (solid
circles) as a function of
temperature.}\label{YbGaGe_thermal_expansion}
\end{center}
\end{figure}

Data taken at subsequent temperatures were used to determine the
temperature dependence of the lattice parameters. Generally, the
lattice parameters we have found at room temperature, compare very
well to those reported elsewhere\cite{Salvador,Muro04,Bobev04}. We
find no evidence of an anomalous thermal expansion, as can be seen
in Figure~\ref{YbGaGe_thermal_expansion}. In fact, we find that
both lattice parameters $a$ and $c$ in our sample increase
smoothly with increasing temperature in the range 20 -- 300~K (see
inset of Figure~\ref{YbGaGe_thermal_expansion}).  In addition, our
estimate of the room temperature volume expansion coefficient
$\beta(300~K)\approx 4.35 \times 10^5$~K$^{-1}$ is comparable to
other metals such as Cu ($\sim 5\times 10^5$~K$^{-1}$)~\cite{Nix}.
This is in contrast to the zero thermal expansion reported by
Salvador et al.~\cite{Salvador}. The normal thermal expansion we
have found is in agreement with the work by Muro et
al.\cite{Muro04}, and by Bobev et al.\cite{Bobev04}.

\section{Conclusions}

The reported crystal structure of YbGaGe\cite{Salvador}, including
the site occupations of Ga and Ge, is confirmed by our neutron
diffraction experiments. In agreement with Refs. \cite{Bobev04}
and \cite{Muro04}, zero or negative thermal expansion is not
observed in our well-characterized bulk sample of YbGaGe, instead
the thermal expansion is comparable to that of Cu ($\sim 5 \times
10^{-5}$~K$^{-1}$)\cite{Nix}. Furthermore, from bulk magnetization
results there is no evidence for a valence change at temperatures
down to 1.8~K, and the small, but non-zero signal is attributed to
about 0.5 wt\% of Yb$_2$O$_3$. The presence of Yb$_2$O$_3$ in our
sample was confirmed by neutron diffraction.

In addition to a normal thermal expansion, differential thermal
analysis performed in an atmosphere with an extremely low partial
pressure of oxygen, and using an inert Ta crucible, indicated our
sample of YbGaGe to melt congruently at about 920~$^\circ$~C. This
is in sharp contrast to results reported in Ref.~\cite{Salvador}.
The elemental analysis reported by Salvador et al. and the atomic
emission spectroscopy on our sample agree very well. Since the
liquidus should not have a discontinuous peak at the melting point
of a compound~\cite{Okamoto91}, samples with similar composition
should have similar melting temperatures. Therefore, we find it
difficult to reconcile our differential thermal analysis with the
analysis reported, since both samples have a very similar
composition, but behave very differently when heated.

We think that the differences between our results and those of
Salvador et al. may be related to different purities of the
starting elements or due to different crucible materials.

\section{Acknowledgments}

The authors wish to thank M. Angst and S.L. Bud'ko for their invaluable input
during the preparation of this manuscript, and A.C. Lawson, J. Zarestky and
V.O. Garlea for their
assistance with neutron data. Ames Laboratory is operated for the
US Department of Energy by Iowa State University under Contract
No. W-7405-Eng-82. This work has benefited from the use of the Los
Alamos Neutron Science Center at the Los Alamos National Laboratory.
This facility is funded by the US Department of Energy under Contract
W-7405-Eng-36. This work was supported by the Director for
Energy Research, Office of Basic Energy Sciences.


\begin{thebibliography}{99}
\bibitem{Salvador} J.~R. Salvador, F. Guo, T. Hogan, and M.G.
  Kanatzidis, \emph{Nature} \textbf{425}, 702 (2003), and J.~R.
  Salvador, F. Guo, T. Hogan, and M.G.
  Kanatzidis, \emph{Nature} \textbf{426}, 584 (2003).

\bibitem{Margadonna04} S. Margadonna, K. Prassides, A.N. Fitch, J.R.
  Salvador, and M. G. Kanatzidis, J. Am. Chem. Soc. \textbf{126}, 4498
  (2004).

\bibitem{[ref1]} A. W. Sleight, \emph{Inorg. Chem.} \textbf{37}, 2854 (1998).
\bibitem{[ref2]} A. C. Lawson, J. A. Roberts, B. Martinez, J. W.
Richardson, Jr., \emph{Phil. Mag. B} \textbf{82}, 1837 (2002); S.
Y. Savrasov, G. Kotliar, E. Abrahams, \emph{Nature (London)}
\textbf{410}, 793 (2002).

\bibitem{Li94} H. Li, C.Y. Wu, and J.C. Ho, \emph{Phys. Rev. B}
  \textbf{49}, 1447 (1994).

\bibitem{Bobev04} S. Bobev, D.J. Williams, J.D. Thompson, and J.L.
Sarrao, \emph{Solid State Commun.}, in print (cond-mat0405063)

\bibitem{Muro04} Y. Muro, T. Nakagawa, K. Umeo, M. Itoh, T.
Suzuki, and T. Takabatake, \emph{J. Phys. Soc. Jpn}, in print

\bibitem{GSAS} A.~C. Larson and R.~B. Von Dreele, \emph{GSAS --
    General Structure Analysis System}, Los Alamos National Laboratory
  Report No. LA-UR-86-748.

\bibitem{Nix} F. C. Nix and D. MacNair, \emph{Phys. Rev.} \textbf{60},
  597 (1941)

\bibitem{Okamoto91} H. Okamoto and T.B. Massalski, \emph{J. Phase
Equilibiria} \textbf{12}, 148 (1991)

\end{thebibliography}
\end{document}